\documentclass[aps,prd,amsmath,floats,floatfix, twocolumn,
superscriptaddress,
  nofootinbib,showpacs]{revtex4} 
\usepackage{graphicx} 
\usepackage{bm}

\usepackage{ulem} 
\usepackage[usenames]{color}

\newcommand{\Caltech}{\affiliation{Theoretical Astrophysics 350-17,
    California Institute of Technology, Pasadena, CA 91125}}
\newcommand{\Cornell}{\affiliation{Center for Radiophysics and Space
    Research, Cornell University, Ithaca, New York, 14853}}
\newcommand{\CITA}{\affiliation{Canadian Institute for Theoretical
    Astrophysics, 60 St.~George Street, University of Toronto,
    Toronto, ON M5S 3H8, Canada}} %

\begin{document}

\title{Measuring orbital eccentricity and periastron advance in quasi-circular black hole simulations.}

\author{Abdul H. Mrou\'e} \Cornell \CITA
\author{Harald P. Pfeiffer} \Caltech \CITA
\author{Lawrence E. Kidder} \Cornell 
\author{Saul A. Teukolsky} \Cornell \Caltech

\date{\today}

\begin{abstract}
  We compare different methods of computing the orbital eccentricity of
  quasi-circular binary black hole systems using the orbital variables
  and gravitational wave phase and frequency. For eccentricities of 
  about a per cent, most methods work satisfactorily.  For
  small eccentricity, however, the gravitational wave phase allows a
  particularly clean and reliable measurement of the eccentricity.
  Furthermore, we measure the decay of the orbital eccentricity 
  during the inspiral and find reasonable agreement with post-Newtonian 
  results. Finally, we measure the periastron advance of non-spinning 
  binary black holes, and we compare them to post-Newtonian 
  approximations.  With the low uncertainty in the measurement 
  of the periastron advance, we positively detect deviations 
  between fully numerical simulations and post-Newtonian calculations.
\end{abstract}


\maketitle


\section{Introduction}



The inspiral and merger of binary black holes or neutron stars is one of 
the most promising sources for current and future generations of
gravitational wave detectors such as LIGO and VIRGO.
The late stage of the inspiral, corresponding to the final few orbits 
and merger of the binary, is highly dynamical and involves strong 
gravitational fields, and it must be handled by numerical relativity. 
Breakthroughs in numerical relativity have allowed a system of two 
inspiraling black holes 
to be evolved through merger and the ringdown of the remnant black hole
~\cite{Pretorius2005a,Pretorius2006,Campanelli2006a,Baker2006a,Campanelli-Lousto-Zlochower:2006,Herrmann2007b,Diener2006,Scheel2006,Sperhake2006,Bruegmann2006,Marronetti2007,Etienne2007,Szilagyi2007,Boyle2007,Scheel2009}.

During the inspiral of an isolated binary, the orbit circularizes via 
the emission of gravitational waves~\cite{Peters1964,PetersMathews1963}.
 As a result, even binaries starting with some eccentricity at the 
beginning of their stellar evolution are expected to have negligible 
eccentricity by the time the frequency of the emitted gravitational 
radiation enters the frequency band of ground based detectors.

However, different physical scenarios~\cite{Kozai1962,Gultekin-Miller-Hamilton2003,Miller-Hamilton2002,Chaurasia-Bailes2005,Ford-Kozinsky-Rasio2000,Ford-Kozinsky-Rasio2004,Wen2003,Melvyn2005,OLeary2009} suggest that 
binaries could approach merger with a significant eccentricity
without being circularized by radiation reaction. This implies that 
eccentric binaries are a potential gravitational wave source for ground 
based interferometers. For example, in globular clusters, the Kozai 
mechanism~\cite{Kozai1962} could increase the eccentricity of an inner 
binary's orbit through a secular resonance caused by a third perturbing
 black hole on an outer orbit~\cite{Miller-Hamilton2002}. Many-body 
encounters of black holes in globular clusters could also result in the 
merger of highly eccentric binaries~\cite{Gultekin-Miller-Hamilton2003}.
 Ref.~\cite{Wen2003} predicted that 30\% of the hierarchical triple 
black hole systems formed in a globular cluster will possess 
eccentricities  greater than 0.1 when their emitted gravitational waves 
pass through a frequency of 10Hz.    

For these reasons considerable attention has been paid to eccentric 
binaries.  Analytical waveform templates have been constructed for the 
gravitational wave signal emitted by compact binaries moving in 
inspiraling eccentric orbits~\cite{Damour2004,KonigsdorfferGopakumar2006,Memmesheimer-etal:2004}.
In this case, orbits involve three different time scales: orbital 
period, periastron advance and radiation reaction time scales. By 
combining these three time scales, one computes ``postadiabatic'' 
short-period contributions to the orbital phasing and gravitational wave
 polarizations. These gravitational wave polarizations are needed for
 astrophysical measurements with gravitational wave interferometers. 
 Refs.~\cite{MartelPoisson1999,Cokelaer2009,BrownZimmerman2009} 
investigated the impact of eccentricity on gravitational wave detection,
 specifically the potential loss in the signal-to-noise ratio when 
``circular'' waveform templates are applied to search for eccentric 
binaries. 

 Eccentric black hole binaries have also been studied with direct 
numerical simulations. Ref.~\cite{Birjoo2009} studied the variation of
 the signal to noise of the eccentric evolutions of intermediate mass 
binary black hole mergers as a function of mass and eccentricity.
Ref.~\cite{HealyEtAl2009} presented binary black holes in zoom-whirl 
orbits where the waveforms are modulated by the harmonics of these 
zoom-whirls. In Ref.~\cite{Sperhake2008Ecc}, the authors studied the
 transition from inspiral to plunge in general relativity by computing 
gravitational waveforms of eccentric nonspinning, equal mass black-hole
 binaries. They analyzed the radiation of energy and angular momentum in
 gravitational waves, the contribution of different multipolar 
components and the final spin of the remnant black hole. 
Ref.~\cite{HinderBirjoo2008} presented results from numerical 
simulations of equal-mass, nonspinning binary black hole inspiral and 
merger for various eccentricities, and they measured the final mass and
 spin of the remnant black hole. Ref.~\cite{HinderEcc2008} compared a 
numerical relativity simulation of an eccentric binary system with 
eccentricity $0.1$ with corresponding post-Newtonian (PN) results. 
They found better agreement when the eccentric PN 
expressions are expanded in terms of the frequency-related parameter 
$x\equiv(\Omega M)^{2/3}$, where $\Omega$ is orbital frequency and 
$M$ is total mass of the binary, rather
than the mean motion $n=2\pi/P$, where $P$ is the orbital period.

Beyond the Newtonian limit, the orbital eccentricity is not uniquely 
defined and a variety of definitions have appeared in the literature.
Ref.~\cite{Lincoln-Will:1990} used a definition of the eccentricity for
which a Newtonian orbit is momentarily tangent to the true orbit (the 
``osculating'' eccentricity), while other authors~\cite{Damour-Schafer:1988,Damour2004,KonigsdorfferGopakumar2006,Memmesheimer-etal:2004} defined
multiple ``eccentricities'' to encapsulate different aspects of 
noncircular orbits at PN order. Another useful definition for large 
eccentricity in numerical simulations is given in 
Refs.~\cite{Berti2006,Will-Mora:2002}.

Similarly, numerical relativists~\cite{Buonanno-Cook-Pretorius:2007,Baker2006d,Pfeiffer-Brown-etal:2007,Husa-Hannam-etal:2007,CampanelliEtal2009}
 introduced several methods for defining and measuring the eccentricity
 using the residual oscillations in the orbital frequency, proper 
horizon separation and coordinate separation. These eccentricity 
definitions are necessary to compare the numerical waveforms with the 
waveforms produced by analytic techniques (i.e., PN methods).
 They behave differently depending on the magnitude of the eccentricity 
and details of the numerical simulation, like employed gauge conditions,
 or presence of numerical noise. This makes it important to specify the
 validity regimes of these definitions.

This paper deals with two related topics: First, we revisit
  many of the eccentricity definitions used so far in numerical work
  and compare them systematically.  We find that for eccentricities of
  a few percent, most definitions work satisfactorily.  However, for
  very small eccentricities, $e\sim 10^{-4}$, computation of the
  eccentricity based on the extracted gravitational waves is superior.
  In the second part of the paper, we measure decay of orbital
  eccentricity and periastron advance for inspiraling black hole
  binaries, and compare these measurements to post-Newtonian 
calculations.  

Section~\ref{sec:eccentricity-estimator} summarizes eccentricity 
definitions that are useful for measuring eccentricity in quasi-circular
 runs. In Section~\ref{sec:eccentricity-estimator}, we compare these 
approaches, as well as some new ones, on the 15-orbit inspiral presented
 by Boyle et al.~\cite{Boyle2007} and on the data of a new simulation of
 an eccentric ($e=0.05$) nonspinning equal mass binary black hole. 
Next, by measuring the extrema in the eccentricity estimator, we 
estimate in Sec.~\ref{sec:Peter-Mathew} the decay of the eccentricity of
 these runs as well as the radial frequency. This allows us in 
Section~\ref{sec:periastron-advance} to estimate the periastron advance 
for these runs from the ratio of the orbital frequency to the radial 
frequency as well as the periastron advance of a set of quasi-circular
 nonspinning binaries of mass ratios 2, 3, 4 and 6. The numerically 
estimated periastron advance is then compared to the 3PN formula of the
 periastron advance~\cite{Memmesheimer-etal:2004,Damour-Schafer:1988,DJS2000}. 

\section{Eccentricity estimators}
\label{sec:eccentricity-estimator}
\subsection{Definitions}
\label{sec:eccentricity}

For a non-precessing binary in  an orbit with zero eccentricity, orbital
 variables and their time derivatives change monotonically as the holes 
inspiral to merger.  In numerical simulations, however, a small 
eccentricity is introduced by imperfections of the initial data. As a 
result, small residual oscillations with amplitude proportional to the 
eccentricity are added to the monotonically changing orbital variables 
and their derivatives. To estimate the eccentricity, one needs to 
determine these residual oscillations.

Different methods to estimate the eccentricity~\cite{Pfeiffer-Brown-etal:2007,Husa-Hannam-etal:2007,Buonanno-Cook-Pretorius:2007} used the 
orbital frequency, separation between the holes (coordinate or proper 
separation), or some Newtonian formula containing both of these 
variables. Similarly, time derivatives of these variables could be used
 in these definitions of the eccentricity. Basically all approaches 
construct an {\em eccentricity estimator} $e_X(t)$ such that for 
Newtonian orbits 
\begin{equation} \label{eq:e1}  
  e_X(t) = e \cos(\Omega_r t+\phi),      
\end{equation} 
where $e$ is the  
eccentricity\footnote{The eccentricity $e$ is well-defined for Newtonian
 orbits.} and $\Omega_r$ is the frequency of radial oscillations in the
 quasi-circular orbit. The key property of $e_X(t)$ is that it is an 
oscillating function with amplitude equal to $e$.

In order to define eccentricity for general relativistic inspirals, one 
computes a tentative eccentricity estimator $e_X(t)$, and checks its 
behavior.  If it behaves as Eq.~(\ref{eq:e1}), one reads off the 
eccentricity $e$ as the amplitude of the oscillations. The resulting 
eccentricity estimates are not local in time nor continuous functions 
of time but rather orbit-averaged quantities.    
Deviation from sinusoidal behavior indicates that particular eccentricity
 estimator is not reliable, and one must verify to what extent the 
eccentricity estimators behave as expected and to what extent they agree.

The estimated value of the eccentricity will differ slightly depending 
on the method used and the noise in the numerical data. In this paper, 
we compare typical eccentricity estimates using a Newtonian formula as 
in Ref.~\cite{Buonanno-Cook-Pretorius:2007} or the orbital frequency and
 separation as in Ref.~\cite{Husa-Hannam-etal:2007}. These 
eccentricities are also compared to new ones computed from the wave 
phase and frequency extracted at a given radius. Other definitions of 
the eccentricity could be used, but we restrict the study to these 
typical definitions. 

To make this rather abstract discussion more concrete, consider the 
Newtonian formula for the radial distance $d$ between the two objects 
with eccentricity 
$e_{\rm Newt}$
\begin{equation}\label{eq:da}
    d(t) = d_0 \left[ 1+ e_{\rm Newt} \cos(\Omega_r t+\phi_0) \right] + O(e^2)\, .
\end{equation}
Based on this formula, one can define the eccentricity estimator $e_{d}(t)$
\begin{equation}
   e_{d}(t) \equiv \frac{d(t) - \overline{d}(t)}{\overline{d}(t)} = e \cos(\Omega_r t+\phi_0), 
\end{equation} 
where the average distance $\bar{d}$ equals $d_0$ in Newtonian gravity. 
For a general relativistic system, one obtains $\bar d(t)$ by a fit over
 several radial oscillation periods.  If the residual $d(t)-\bar d(t)$ 
oscillates sinusoidally---which it indeed does for sufficiently large 
eccentricity---the amplitude of these oscillations defines an associated
 eccentricity $e_d$.

From the trajectory of the two objects, one can also use the orbital 
phase and frequency to define the corresponding eccentricity estimators
 using the following Newtonian relation~\cite{Murray-Dermott}:
\begin{equation}
\label{eq:fm}
\Phi={\cal M} + 2e \sin {\cal M } + \frac{5}{4} e^2 \sin2 {\cal M}+ O(e^3)\,,
\end{equation}
where $\cal M$ is the mean anomaly and $\Phi$ is the orbital phase. 
Equivalent to Eq.~(\ref{eq:fm}) for numerical simulations is the 
relationship
\begin{equation}
\label{eq:Phi}
\Phi(t)= \Phi_0 + \Omega_0 t + 2 e \sin (\Omega_r t ) + O(e^2)\,,
\end{equation}
where $\Omega_0$ is the average fitted orbital frequency and $\Phi_0$ is
 some phase offset. Then the eccentricity estimator $e_\Phi(t)$ is 
written as 
\begin{equation}
\label{eq:eop}
e_\Phi(t)= \frac{ \Phi(t) - \Phi_0 - \Omega_0 t }{ 2 }\,.
\end{equation}
From the time derivative of Eq.~(\ref{eq:Phi})  and the replacement 
$\Omega_r\to\Omega_0$, we obtain an eccentricity estimator in terms of 
the orbital frequency 
(as in Ref.~\cite{Husa-Hannam-etal:2007})
\begin{equation}
\label{eq:eof}
e_\Omega(t)= \frac{\Omega(t)-\Omega_0}{2\Omega_0} \,. 
\end{equation}

Notice that since the radial oscillation results from eccentricity, 
$\Omega_r$ is different from $\Omega_0$, the average of the orbital 
frequency. The eccentricities of Eqs.~(\ref{eq:eop}) and~(\ref{eq:eof}) 
will differ by a factor $\Omega_r/\Omega_0$.  For Newtonian orbits, 
$\Omega_r/\Omega_0 = 1$, and this factor drops out.  But for the binary
 black hole case, the factor is about 1.4, 
 causing the difference between Figs.~\ref{fig:FilteredWavePhase} and 
~\ref{fig:FilteredWaveFrequency} below. This is easily seen by writing 
the eccentricity estimator from Eqs.~(\ref{eq:Phi}) and~(\ref{eq:eof}) as
\begin{equation}
   \frac{\dot{\Phi}(t) - \Omega_0}{2 \Omega_0}  = e \Omega_r/\Omega_0 \sin(\Omega_r t)\,.
\end{equation}

\subsection{Numerical data}

Before introducing several further eccentricity estimators, let us 
briefly describe the numerical binary black hole simulations that we 
will analyze.  All runs have been performed with the Spectral Einstein Code (SpEC)~\cite{SpECwebsite}.  We will primarily analyze the 16 orbit long inspiral 
simulation of an equal mass, non-spinning black hole binary presented in
 Ref.~\cite{Boyle2007} (specifically, the run labeled 30c-1).  This run
  with eccentricity of about $6\times 10^{-5}$ is used to compute the 
eccentricity data in Figs.~\ref{fig:BCP}, \ref{fig:Jena}, 
\ref{fig:FilteredWavePhase} and \ref{fig:FilteredWaveFrequency}.
 To compute eccentricity estimators, we use the orbital frequency 
$\Omega$, the coordinate separation between the holes $D$, the proper 
horizon separation $s$  (defined as the integrated distance between the 
holes along the coordinate axis, cf. Ref.~\cite{Boyle2007}) as well as 
the gravitational wave phase $\phi$ and the gravitational wave frequency
 $\omega$.
 
\begin{figure}
\includegraphics[scale=0.47]{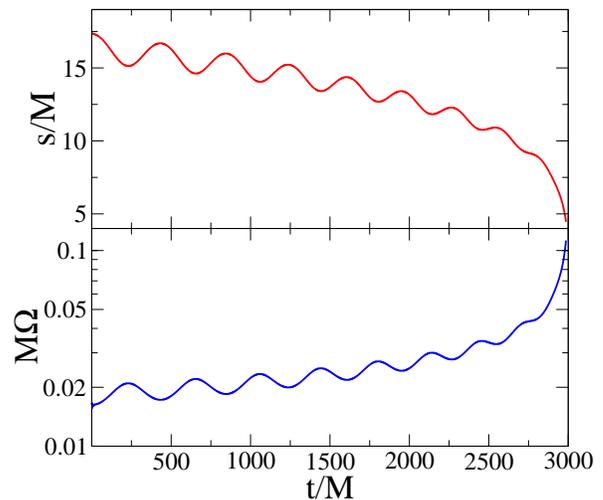}
\caption{\label{fig:PropSepe05} The equal mass nonspinning binary run 
with eccentricity $e\sim0.05$. As a function of time, the top panel 
shows the proper horizon separation and the bottom panel shows the 
orbital frequency. For such a value of the eccentricity,
it is easy to measure the decay rate of the eccentricity and
estimate the periastron advance of the binary near the merger.}
\end{figure}

 Furthermore, we utilize recent runs of quasi-circular nonspinning 
binaries~\cite{Buchman-etal-in-prep}   with mass ratios 2, 3, 4 (lasting
 15 orbits) and mass ratio 6 (lasting 8 orbits). The eccentricity of 
these runs is also of the order of magnitude $10^{-5}$. The periastron 
advance and the resulting frequency modulation are estimated in 
Fig.~\ref{fig:precession-qneq1}.

As a separate check, another equal mass nonspinning binary with moderate
eccentricity ($e\sim0.05$) is evolved to compare various eccentricity 
estimators and measure the periastron advance for a case that is not 
quasi-circular. Figure~\ref{fig:PropSepe05} shows the proper 
separation as well as the orbital frequency as a function of time for 
this eccentric binary.

\subsection{A Newtonian definition}

The first use of eccentricity estimators was by Buonnano, Cook 
\& Pretorius~\cite{Buonanno-Cook-Pretorius:2007}, who consider 
the following relationship that holds
 for Newtonian orbits with eccentricity $e_{\rm Newt}$:
\begin{equation} \label{eq:Omega}
  \left[\Omega_\phi(t)^2 r(t)^3/M-1\right]  = e_{\rm Newt} \cos \phi(t)\,.
\end{equation}
Here $\Omega_\phi(t)$ and $\phi(t)$ denote orbital frequency and
  phase, respectively, and $r$ is the separation of the masses.
Motivated by Eq.~(\ref{eq:Omega}), 
Buonnano, Cook \& Pretorius define an eccentricity estimator
\begin{equation} \label{eq:DOmega}
 e_{\rm BCP}(t) = \Omega_\phi(t)^2 r(t)^3/M - \left[\Omega_\phi(t)^2 r(t)^3/M\right]_{\rm fit}\,,
\end{equation}
where now $\Omega_\phi(t)$ and $r(t)$ are extracted from the
  numerical simulation.
To compute this eccentricity estimator $e_{\rm BCP}$, we fit the function 
$\Omega_\phi(t)^2 r(t)^3/M$ to a polynomial in time,
\begin{eqnarray} \label{eq:f}
  f(t) = \sum_{i=0}^{n} a_i t^{i}\,.
\end{eqnarray}
We found that a fifth order polynomial ensures a good fit. 
The polynomial order needs to be high enough to reliably capture the 
smooth inspiral trend in $\Omega(t)^2 r(t)^3/M$, but it should {\em not}
 capture the higher frequency oscillations due to eccentricity.  When 
applying this procedure to a binary black hole inspiral, one has to 
decide how to generalize the Newtonian separation $r(t)$ to curved 
space.  We use two choices, the coordinate distance $D(t)$ between the 
centers of the apparent horizons, and the proper separation $s(t)$ 
between the apparent horizons, computed along a straight coordinate line
 connecting the centers of the apparent horizons.  

\begin{figure}
\includegraphics[scale=0.47]{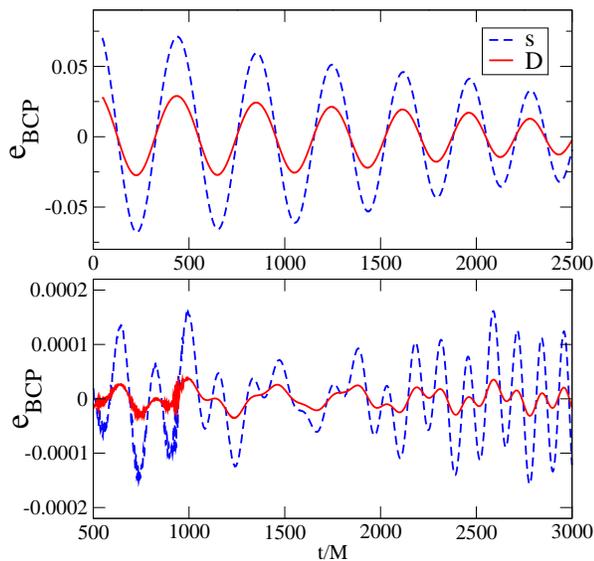}
\caption{\label{fig:BCP} Eccentricity estimator $e_{\rm BCP}$~\cite{Buonanno-Cook-Pretorius:2007} applied to a simulation with $e\sim 0.05$ (top panel) and  $e\sim 6\times 10^{-5}$ (bottom panel). The dashed and solid lines correspond to $e_{\rm BCP}(t)$ computed from the coordinate 
separation and the proper horizon separation. For the large eccentricity run, $e_{\rm BCP}$ exhibits
clear oscillations, whereas for the small eccentricity run, $e_{\rm BCP}$ is
dominated by other features.  In both cases, the amplitude of $e_{\rm BCP}$ 
is smaller when defined using coordinate distance $D$. }
\end{figure}

In Fig.~\ref{fig:BCP}, we plot the eccentricity estimator $e_{\rm BCP}$ computed 
using the coordinate separation and proper horizon separation as 
described above. 
In the top panel, we plot $e_{\rm BCP}(t)$ using the binary 
run with eccentricity $e\sim0.05$. 
Using the proper horizon separation $s$, the estimated initial eccentricity, 
$0.07$, is larger by nearly a factor of 2 than in 
the case where the coordinate separation $D$ is used ($0.03$).
 This is due to different numerical values for the distances, 
$(s/D)^3\sim 1.8$. Both 
eccentricity estimators are in phase during the whole time interval as 
expected. In both cases, the eccentricity magnitude decreases between $t=0$ 
and $t=2500M$. In this case, 
a clear decaying sinusoidal signal is obtained without any higher harmonics 
showing up at later times.

In the bottom panel, we examine the equal mass binary with eccentricity 
$e\sim6\times 10^{-5}$.
For this case, no clean sinusoidal signal is apparent.  While
  $e_{\rm BCP}$ computed from $s(t)$ shows oscillations, they are {\em
    faster} than the orbital period, and can therefore not be
  attributed to orbital eccentricity.  Because $e_{\rm BCP}$ does not
  show the expected behavior, it is not meaningful to attribute a
  value of eccentricity to this analysis.  For these small
  eccentricities, $e_{\rm BCP}$ is dominated by other effects, possibly the
  coordinate dependence of the separation measurements.

\subsection{ Eccentricity from orbital variables}

Husa et al~\cite{Husa-Hannam-etal:2007}
 fitted directly
the orbital frequency $\Omega(t)$ or the coordinate separation $D(t)$
to a function of the form 
\begin{eqnarray} \label{eq:D}
  X_{\rm fit}(t) = \sum_{i=1}^{n} a_i (t_m-t)^{i/2}\,,
\end{eqnarray}
with fitting parameters $t_m$, the coalescence time, and the
coefficients $a_i$. 
 The eccentricity estimator is then defined as
\begin{equation} \label{eq:eD}
  e_{\rm X}(t) = \frac{X_{\rm NR}(t)-X_{\rm fit}(t)}{k X_{\rm fit}(t)}\,,      
\end{equation}
where $X_{\rm NR}(t)$ is the numerical orbital variable
and $X_{\rm fit}(t)$ is the polynomial fit of $X_{\rm NR}(t)$. 
We shall compute three eccentricity estimators using Eq.~(\ref{eq:D}), which differ in the
quantity being fitted:  $e_s(t)$ and $e_D(t)$ are based on proper separation 
and coordinate separation between the black holes, with the value $k=1$; 
$e_\Omega(t)$ uses the orbital frequency, where $k=2$. 
In the Newtonian limit, these estimators are identical to first order in 
eccentricity.

\begin{figure}
\includegraphics[scale=0.5]{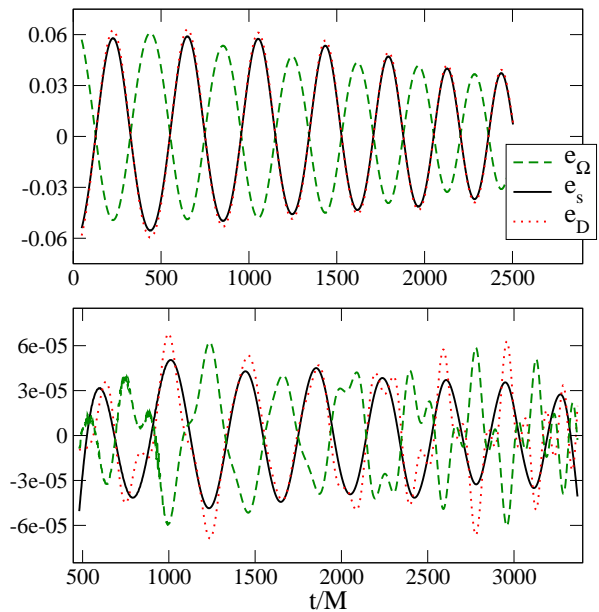}
\caption{\label{fig:Jena}  
Eccentricity estimators based on orbital trajectories applied
to simulations with eccentricity $e\sim 0.05$ (upper panel) and
$e\sim 6\times 10^{-5}$ (lower panel). The quantities
$e_\Omega$, $e_{s}$ and $e_{D}$ are computed from orbital
 frequency, proper horizon separation and coordinate separation using 
Eq.(\ref{eq:eD}).} 
\end{figure}

Figure~\ref{fig:Jena} shows these eccentricity estimators for a
run with fairly large eccentricity and for a run with very small eccentricity.
 For large eccentricity $e=0.05$, the various eccentricity estimators have a 
smooth decaying sinusoidal
signal. This allows measuring a nearly identical value of the eccentricity 
for the three orbital variables from the amplitude of the residual oscillations.
The phasing is also consistent between the different eccentricity estimates:
The orbital frequency is a maximum when the separation is a minimum and 
vice-versa.  

In the bottom panel of Fig.~\ref{fig:Jena}, we plot the eccentricity estimators
 applied to a simulation with much smaller eccentricity $e\sim 6\times10^{-5}$.
The behavior of $e_D$ and $e_\Omega$ is erratic.  Higher-order
harmonics are clearly visible, and the extrema are not monotonically
decreasing, as one would expect from the circularizing effect of
gravitational radiation.
 However, $e_s$ shows no increase in the eccentricity 
 during the late stages of the inspiral, and no additional significant harmonics
appears even at $t=3500M$. 
The order of the polynomial fit depends on the time range of the fit. In this 
case, a fifth order polynomial was enough to capture the oscillatory 
behavior in the eccentricity estimator in the time range $500M<t<3500M$.
Note that the orbital phase could also be used to measure the  
eccentricity estimator using Eq.(\ref{eq:eD}) (but without division
by $X_{\rm fit}$).

\subsection{Eccentricity from gravitational waves}

All eccentricity estimators discussed so far utilize
coordinate-dependent quantities like separation or orbital frequency.
Therefore, one might suspect that the higher harmonics visible in
Figs.~\ref{fig:BCP} and~\ref{fig:Jena}  are caused by
gauge effects.  The gravitational radiation at future null infinity is
expected to be gauge-invariant, removing the dependence on
gauge-dependent quantities. These considerations motivate the use of the
 gravitational wave phase and frequency to define eccentricity.

We extract the $(l,m)=(2,2)$ mode of the gravitational wave using the 
Newman-Penrose scalar $\Psi_4$ and define the wave phase $\phi(t)$ as~\cite{Boyle2007}
\begin{equation}
\Psi_4^{22}(r,t) = A(r,t) e^{-i\phi(r,t)}.
\end{equation}
Then the gravitational-wave frequency is defined as 
\begin{equation}
\omega  = \frac{d\phi}{dt}.
\end{equation}
The waveforms extracted at finite radii are extrapolated to null infinity 
using the procedure in~\cite{Boyle-Mroue:2008}.  The wave phase $\phi$ and 
frequency $\omega$ are measured as a function of the retarded time 
$t-r^*$, where $r^*$ is the tortoise-coordinate radius defined as
\begin{equation}
  r^*\equiv r + 2 M_{\rm ADM} \ln \left( \frac{r}{2 M_{\rm ADM}}-1\right) \ , 
\end{equation}
where $M_{\rm ADM}$ is the ADM mass of the initial data.
At early times, the gravitational waveforms are contaminated by high 
frequency noise from imperfect initial data. 
To measure the amplitudes and locations of the extrema in the  
eccentricity estimator more accurately, the residual functions are filtered 
using a low-pass Butterworth filter with the Matlab function ``filtfilt''
~\cite{Boyle-Mroue:2008}. The filtered data can be used to measure the 
eccentricity for retarded time $t-r^*\gtrsim 1000M$.

Based on the gravitational wave phase, we define the eccentricity estimator
\begin{equation}
\label{eq:ephase}
e_{\phi}(t)=\frac{\phi_{\rm NR}(t)-\phi_{\rm fit}(t)}{4 }\,, 
\end{equation}
where an additional factor of 1/2 arises because the wave phase is 
approximately twice the orbital phase.

\begin{figure}
\includegraphics[scale=0.47]{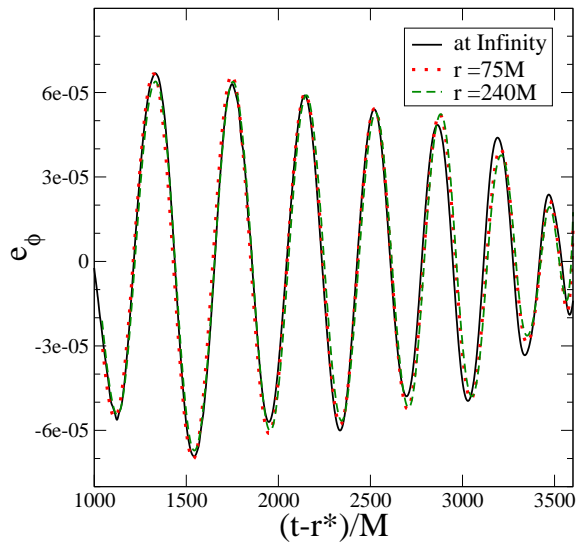}
\caption{
\label{fig:FilteredWavePhase}  Eccentricity estimator $e_\phi$  
computed from the gravitational wave phase as a function of retarded time 
$t-r^*$. In this plot, the eccentricity estimator is computed from the 
gravitational wave extracted at finite radii $r=75M$ and $r=240M$ and from
 data extrapolated to infinity. The three curves agree in amplitude and phase 
to within 5\% in the retarded time interval $1000M<t-r^*<3000M$.}
\end{figure}

In Fig.~\ref{fig:FilteredWavePhase}, we plot the eccentricity estimator computed
from the gravitational wave phase of the (2,2) mode extracted at the radii 
$r=75M$, $r=240M$ and extrapolated to infinity using terms up to $1/r^2$
versus $t-r^*$. The eccentricity estimate is independent of the radius 
value at which the wave is extracted, and various estimates  
agree to within 5\% in both amplitude and phase for different radii of extraction.

\begin{figure}
\includegraphics[scale=0.47]{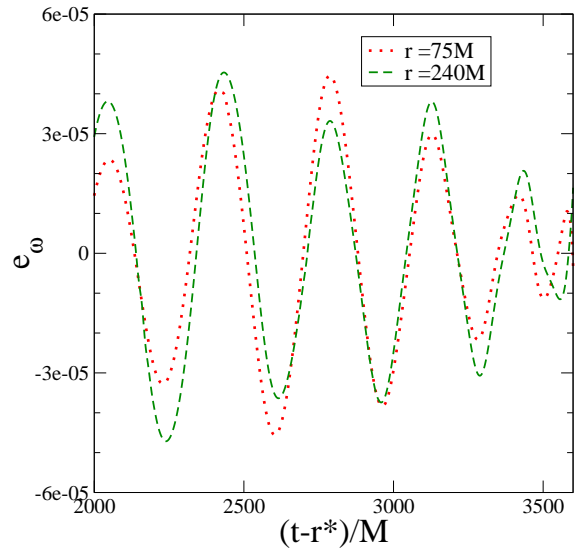}
\caption{\label{fig:FilteredWaveFrequency} 
Eccentricity estimator $e_\omega$ computed from the gravitational wave 
frequency as a function of the retarded time 
$t-r^*$. In this plot, the eccentricity estimator is computed from the 
gravitational wave extracted at $r=75M$ and $r=240M$. The eccentricity 
estimator is contaminated by significant noise caused by imperfect initial 
data at a time earlier than $t/M=2000$. 
}
\end{figure}

Using the wave frequency we define the eccentricity estimator $e_\omega(t)$ 
\begin{equation}
\label{eq:efrequency}
e_{\omega}(t)=\frac{\omega_{\rm NR}(t)-\omega_{\rm fit}(t)}{2 \omega_{\rm fit}(t) }\,.
\end{equation}
Computation of the gravitational wave frequency $\omega=d\phi/dt$ requires a 
derivative of $\phi(t)$, which increases numerical noise.  Given the small 
amplitude of the effect under consideration (the fractional change in $\omega$
 is $2e={\cal O}( 10^{-4})$), the increased noise noticably affects $e_\omega$.
It is usable only at finite extraction radius, and even there only for 
 $t-r^*\gtrsim  2000M$.  

In Fig.~\ref{fig:FilteredWaveFrequency}, we compute the eccentricity estimator 
from the wave frequency extracted at $r=75M$ 
and
$r=240M$. The extrapolated data to infinity is not shown because of its 
sensitivity to noise.
The two curves have a nearly sinusoidal behavior with 
the phase agreeing to within 10\%. However, the amplitude differs by 25\% 
between the wave data measured at $r=75M$ and $r=240M$. The reduced sensitivity
 to noise is an important advantage of $e_\phi$ over $e_\omega$.

For the binary with eccentricity $0.05$, plots similar to 
Figs.~\ref{fig:FilteredWavePhase} and~\ref{fig:FilteredWaveFrequency} with 
smooth sinusoidal behavior could easily be obtained. 

Computation of the eccentricity from gravitational radiation ($e_\phi$ and
  $e_\omega$) is better behaved than the methods using orbital variables. Only 
one harmonic mode appears in the data---even for the low-eccentricity run with $e\sim 6\times 10^{-5}$---and the eccentricity is decreasing
as the binaries inspiral toward merger. We attribute this improvement to
the disappearance of coordinate and gauge effects when the data are extracted 
further away from the holes.

The eccentricities extracted from $e_\phi$ and $e_\omega$ in 
Figs.~\ref{fig:FilteredWavePhase} and ~\ref{fig:FilteredWaveFrequency} are 
inconsistent with each other; they differ by a factor $\Omega_r/\Omega_\phi$ 
as explained in Sec.~\ref{sec:eccentricity}. 

One might also consider a definition of the
 eccentricity based on taking the time derivative of the wave frequency. From
Eq.~(\ref{eq:fm}), the second time 
derivative of the orbital phase is given by:
\begin{equation}
\ddot \Phi = \ddot {\cal M} -2e (\ddot {\cal M} \cos {\cal M} + {\dot{\cal M}}^2  \sin {\cal M} ) +O(e^2)\,,
\end{equation} 
where the amplitude of the oscillatory part is 
$2e\sqrt{\ddot{ \cal M}^2+ \dot {\cal M}^4}$.
Then, the eccentricity estimator computed from the time derivative of the wave frequency 
$e_{d\omega}$ is then defined as
\begin{equation}
e_{d\omega}= -\frac{{\ddot \phi}_{\rm NR}-{\ddot \phi}_{\rm fit}}{2\sqrt{{\ddot \phi}^2_{\rm fit}+ {\dot \phi}^4_{\rm fit}}}\,.
\end{equation} 
The main advantage of a such a definition is that it requires a lower
order fitting polynomial. Unfortunately, the numerical derivatives necessary to
  compute $\ddot{\phi}$ amplify noise, and so this method becomes
  impractical for the numerical evolutions considered.

\begin{table*}
\begin{tabular}{c|c|c|c|c|c|c|c|c|c}
Method & Ecc. Res.& Definition & $t_i/M$ & $t_f/M$  & $n$ & $e(t/M=1000)$ & $e(t/M=2000)$ & $e(t/M=3000)$ & $\delta e/e$
\\\hline
GW Phase & $e_{\phi}$ & $ \Delta \phi/4 $ & 952 & 3861 &   $7$ &  6.4$\times 10^{-5}$ & 5.7$\times 10^{-5}$ & 4.8$\times 10^{-5}$ & 5-15\%
\\
GW Frequency & $e_{\omega}$  & $ \Delta \omega/(2\omega_{\rm fit}) $ & 1922& 3861 &  $7$ & - & 4.3 $\times 10^{-5}$& 3.7$\times 10^{-5}$ & 15-25\%
\\
Coordinate distance       &  $e_{D}$  & $\Delta D/D_{\rm fit}$ & 480 & 3367 & $7$ &  6.7$\times 10^{-5}$& 4.9$\times 10^{-5}$ & 6.3$\times 10^{-5}$ & 15-40\%
\\
Proper separation     &  $e_{h}$  & $ \Delta h/h_{\rm fit} $ & 480 & 3367 &  $5$ & 5.0 $\times 10^{-5}$& 3.9 $\times 10^{-5}$& 3.4$\times 10^{-5}$ & 10-20\% 
\\
Orbital frequency & $e_{\Omega}$  & $ \Delta \Omega/(2\Omega_{\rm fit}) $ & 480 & 3367 &  $7$ & 6.2$\times 10^{-5}$ & 4.1$\times 10^{-5}$ & 3.4$\times 10^{-5}$ & 20-30\%
\\\hline
BCP       &  $e_{\rm BCP}$  & $\Delta(\Omega(t)^2 r^3 )$ & 480& 3367 & $5$ & 3.5$\times 10^{-5}$ & 2.4$\times 10^{-5}$ & 2 $\times 10^{-5}$& 50-80\%
\end{tabular}
\caption{\label{tab:list-ecc} Summary of the eccentricity measurement methods. 
$t_i$ ($t_f$) is the initial (final) time of fitting. n is the order of the 
best fitting polynomial in the time interval $[t_i/M,t_f/M]$.
$e$ is the eccentricity estimate at the time $t$ with the relative error 
$\delta e/e$.}
\end{table*}

In Table~\ref{tab:list-ecc}, we summarize the eccentricity definitions examined
in this paper, the data range between $t_i/M$ and $t_f/M$ employed in the fits, 
and the order of the fitting polynomial $n$ for the 15-orbits quasi-circular 
nonspinning binary. We also give an estimate of the eccentricity value at
$t/M=1000,2000$ and $3000$ and its estimated error $\delta e/e$ for each 
method.

\section{Behavior of eccentricity during inspiral}
\label{sec:Peter-Mathew}

Radiation reaction reduces eccentricity during the inspiral of
  a binary compact object, as shown by the post-Newtonian calculation
  by Peters~\cite{Peters1964}. Using the quadrupole approximation, Peters
derived the evolution of the orbital
eccentricity during the inspiral caused by the emission of gravitational
waves.
In the limit of small eccentricity, the eccentricity
is related to the semi-major axis $a$ by 
\begin{equation} \label{eq:Peters}
  e \propto  a^{19/12}\,.      
\end{equation}
The first
confirmation of the decay of eccentricity in a fully numerical binary
black hole inspiral was presented by
Pfeiffer~et.~al.~\cite{Pfeiffer-Brown-etal:2007}.  Pfeiffer et.~al.
measured the decay rate of the eccentricity for an equal mass,
nonspinning binary with an eccentricity of about 0.02 during the last five 
orbits of the inspiral. The precise
decay rate depended on the definition of the eccentricity
used. For a definition based on the orbital frequency, good agreement with
Eq.~(\ref{eq:Peters}) was found. 

\begin{figure}
\includegraphics[scale=0.47]{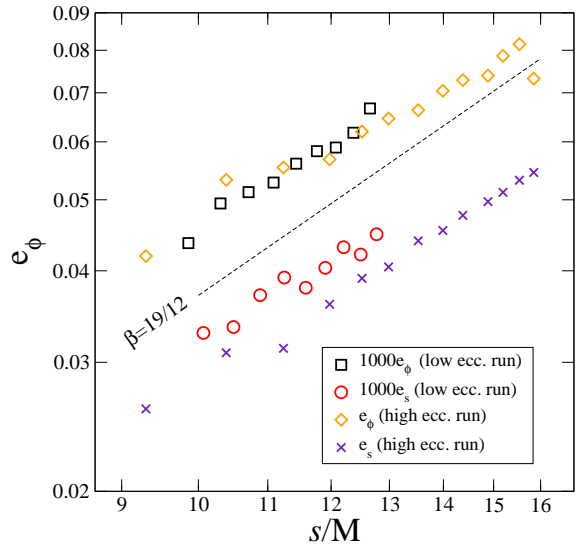}
\caption{\label{fig:ecc-sep2} Eccentricity as a function of proper horizon 
separation. We show data for two simulations, with high and low eccentricity.
 For each run we compute eccentricity from the GW-phase $\phi$ and the proper 
separation $s$. 
The dashed line represents the power-law $s^{19/12}$ predicted by 
post-Newtonian theory (See. Eq.~\ref{eq:Peters}). 
}
\end{figure}

In Sec.~\ref{sec:eccentricity-estimator}, we established that
the eccentricity estimators $e_{\phi}$ (wave phase) and $e_s$ (proper 
horizon separation) show the cleanest oscillatory behavior.  Using these two
 eccentricity estimators, we compute as follows the eccentricity 
 as a function of time for the much longer inspirals considered here.  
 We first define the ``average'' eccentricity over one half of a radial 
oscillation as the difference between two consecutive extrema (from minimum to maximum, or vice versa) of the eccentricity estimator 
\begin{equation}
e= \frac{|A_{\rm min}-A_{\rm max}|}{2}\,.
\end{equation}
We further associate this eccentricity with the time half-way between the two extrema under consideration:
\begin{equation}
t(e)= \frac{t(A_{\rm min})+t(A_{\rm max})}{2}\,.
\end{equation}
At the time of this average eccentricity, the separation is measured 
numerically. In the case when gravitational wave data is used, the wave 
 phase is approximated as a function of the separation by using 
the retarded time $t-r^*$.

The results are plotted in Fig.~\ref{fig:ecc-sep2}.
Fitting a power-law
\begin{equation}
\label{eq:powerlaw}
\log e= \alpha +\beta \log s
\end{equation}
to the numerical data  yields $\beta\approx 1.4$. These decay estimates are in
 reasonable agreement with Peters' prediction ($\beta=19/12\approx 1.583$), as 
can be seen by the indicated power-law in Fig.~\ref{fig:ecc-sep2}. 
The orbital eccentricity decays similarly in the two simulations with different
 eccentricity.

\section{ Periastron advance}
\label{sec:periastron-advance}

The periastron advance is one of the new features for relativistic 
eccentric orbits that is not present in Newtonian gravity. It has been
 computed analytically in the post-Newtonian 
regime up to third order but---to our knowledge---it has never been
estimated numerically in binary black hole simulations. Periastron advance 
will lead to a modulation of the   gravitational wave signal for eccentric 
binaries and will impact gravitational wave detection strategies. Therefore,
 it is important to know what this frequency is and how it changes as a function of the
 mass ratio. The fractional periastron advance per orbit, $K$, is defined as
\begin{equation}\label{eq:DefPeriastronAdvance1}
K \equiv \frac{\Delta\Phi}{2 \pi}\,,
\end{equation}
where $\Delta \Phi=\Phi-2\pi$ is the periastron advance per orbit. 
The dimensionless parameter $K$ is related to the radial frequency $\Omega_r$ 
and the orbital frequency $\Omega_\Phi$ through
\begin{equation}\label{eq:DefPeriastronAdvance2}
\frac{\Omega_\Phi}{\Omega_r}=K+1 \,.
\end{equation}

\subsection{Numerical method for measuring the periastron advance}

From an eccentricity  estimator $e_X$, cf. Eq.~(\ref{eq:eD}), one can read 
off not only the eccentricity (via the amplitude of $e_X$), but also the 
frequency of the radial motion, $\Omega_r$ (from the oscillation period).  
We shall define the period of the radial oscillation as twice the time interval
between two consecutive extrema (from minimum to maximum, or vice versa) in the
eccentricity estimator curve. We employ the following procedure to compute the
periastron advance:
\begin{enumerate}
\item Choose a cleanly oscillating eccentricity estimator $e_X(t)$. We will use $e_{\rm \phi}$, cf. Fig.~\ref{fig:FilteredWavePhase}.
\item Find the extrema of $e_X(t)$. This gives a time list 
($t_0,t_1,...,t_k,...$) corresponding to all 
perihelia or aphelia (i.e., extrema in the residual radial velocity). 
\item Interpolate the orbital phase $\Phi$ to the times $t_k$.
  Between neighboring data points, the orbital phase changes
  by $\Phi(t_{k+1})-\Phi(t_{k-1})$, whereas the radial phase changes by $2\pi$.
  Therefore, the ratio between orbital
  and radial  phase increase is $(\Phi(t_{k+1})-\Phi(t_{k-1}))/2\pi$, and
  so
\begin{equation}
\frac{\Omega_\Phi}{\Omega_r} = \frac{\Phi(t_{k+1})-\Phi(t_{k-1})}{2\pi}.
\end{equation}
\end{enumerate}

For the very
low eccentricity simulation ($e\sim 5\times 10^{-5}$), the periastron advance 
is very difficult to measure because the amplitude of $e_\phi$ is so small. The
uncertainty in the extracted $\Omega_\Phi/\Omega_r$ is about $\pm 0.1$ for $0.02\le M\Omega_\Phi\le0.03$.
 The error in the estimated periastron advance increases at
  higher frequencies as the binary evolves closer to merger. The
eccentricity estimators depend on details of the polynomial fits, and it
is more difficult to read off these small eccentricity residuals near the
plunge. Therefore, $\Omega_\Phi/\Omega_r$ for the $e\sim
  5\times 10^{-5}$ run should not be trusted for $M\Omega_\Phi\gtrsim
  0.03$.

In the simulation with larger eccentricity $e\sim 0.05$, by contrast, the
periastron advance is easier to measure, because the amplitude of
  $e_\Phi$ is proportional to the eccentricity. We obtain correspondingly smaller errors, about 3\% at frequencies $M\Omega\lesssim 0.03$.
While we are able to extract $\Omega_\Phi/\Omega_r$ at higher
  frequencies for the simulation with $e\sim 0.05$, recall that the 
  numerical data are
  constructed from consecutive extrema of $e_\phi$.  At late times
  (close to merger), there is an increasing amount of orbital
  evolution during such an interval, which renders ambiguous
    both the definition of $\Omega_\Phi/\Omega_r$ and its association
    with one orbital frequency.

\begin{figure}
\includegraphics[scale=0.51]{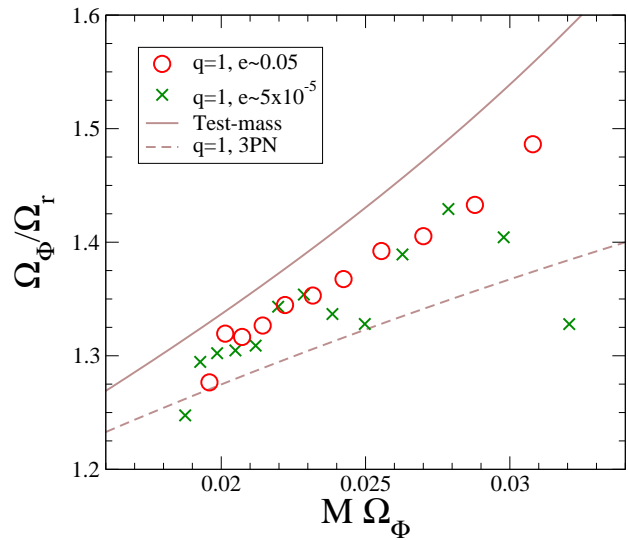}
\caption{\label{fig:precession} Periastron advance for
      equal-mass binaries. Plotted is the ratio of orbital frequency
  to radial frequency, $\Omega_{\Phi} / \Omega_r$, versus
  the orbital frequency $M\Omega_{\Phi}$.
 The data represent numerical simulations of equal mass nonspinning
  black-hole binaries with two different eccentricities
    $e$.  Also shown are the prediction
    of post-Newtonian theory for $q\!\!=\!\!1$ and the test-mass
    result based on geodesic motion in Schwarzschild (both in the
    limit $e\!\ll\!1$). For $e\sim 5\times 10^{-5}$, the numerical data
    is unreliable for $M\Omega_\Phi\gtrsim 0.03$ (see text).}
\end{figure}

Figure~\ref{fig:precession} shows the computed periastron advance 
for the two equal-mass simulations considered here. To
  facilitate comparison with analytical estimates (see next section),
  we plot $\Omega_\Phi/\Omega_r$ as a function of orbital frequency. 
  The latter is approximated as half the gravitational wave frequency.
  (This is justified because the deviation from this
  value is much smaller than
  the error in estimating the eccentricity and the periastron
  advance.) We will discuss this figure in the next subsection.

\subsection{Results}

From Fig.~\ref{fig:precession} we see that 
    $\Omega_\Phi/\Omega_r$ is positive (i.e. the fully general
    relativistic calculation produces indeed a periastron {\em
      advance}), and the periastron advance increases with increasing
    orbital frequency $M\Omega_\Phi$, again consistent with
    expectations.  
The solid and the dashed lines in Fig.~\ref{fig:precession}
  indicate the periastron advance for a test-mass orbiting a
  Schwarzschild black hole, and for an equal-mass binary at 3rd
  post-Newtonian order (see Appendix for details), and we can now
  compare these calculations with the fully relativistic BBH
  simulations.  The scatter in the numerical data
  $\Omega_\Phi/\Omega_r$ represents a measure of the uncertainty in
  the periastron advance of the numerical simulations.  For the $e\sim
  0.05$ simulation, this scatter is much smaller than the difference
  from the 3PN calculation.  Therefore, we have positively detected a
  difference between fully numeric simulations and 3PN calculations.   
  ($\Omega_\Phi/\Omega_r$
  from the $e\sim 10^{-5}$ simulation coincides with the
  data for the $e\sim 0.05$ run, although with larger scatter,
  because of trying to extract much smaller variations in the
  numerical data.)   The
  difference between the numerical periastron advance curve and the 3PN
  result is about 3\% at $\Omega_\Phi=0.02$ and continually increases to 
  about 5\% at $\Omega_\Phi=0.03$.   The fully NR periastron advance
  seems to follow more closely the test-mass calculation than the
  equal-mass 3-PN prediction.  Note that comparing either of the two
  analytic results is imperfect: The 3-PN calculation is for equal
  masses, but because of the nature of post-Newtonian perturbation theory
  becomes increasingly less reliable for increasing frequency
  $M\Omega_\Phi$.  The test-mass limit, in contrast, is an exact
  calculation, but for a system different from an equal-mass
  binary.  Unequal mass binaries with mass ratios very different from
  unity should result in better agreement with the test-mass limit,
  and we will explore this case next.

\begin{figure}
\includegraphics[scale=0.51]{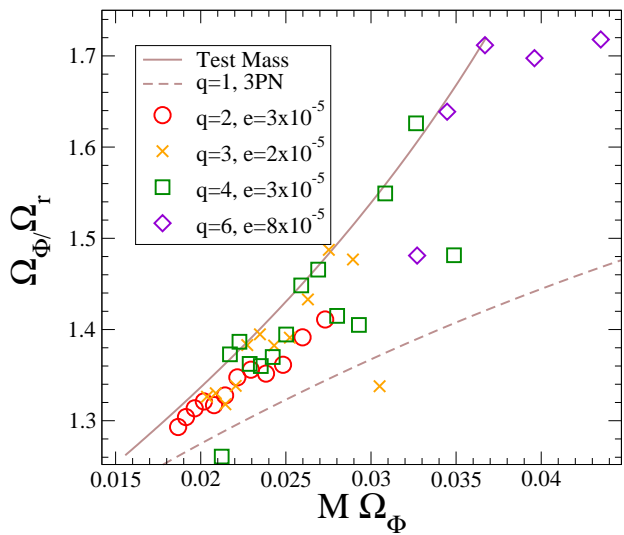}
\caption{\label{fig:precession-qneq1} 
Periastron advance for unequal mass BBH. 
Shown is the ratio of orbital frequency to radial frequency, 
$\Omega_{\Phi} / \Omega_r$, versus the orbital frequency $M\Omega_{\Phi}$
for different mass-ratios $q=M_1/M_2$.}
\end{figure}

Extracting the periastron advance from a series of
  non-spinning unequal mass simulations~\cite{Buchman-etal-in-prep},
  we obtain the data plotted in Fig.~\ref{fig:precession-qneq1}.  
  These simulations have very low eccentricity
  in order to accurately model circularized binaries for gravitational
  wave data-analysis, with eccentricities indicated in
  Fig.~\ref{fig:precession-qneq1}. The smallness of the eccentricity is 
  unfortunate for our purposes, as this increases the errors in the extracted
  periastron advance. The periastron advance for $q=2$ is very similar
  to the equal mass periastron advance data. For higher mass ratio, the 
numerically computed $\Omega_\Phi/\Omega_r$ seems to
  increase and approach the test-mass result; however, the large uncertainty in
  $\Omega_\Phi/\Omega_r$ for these runs prevents us from drawing strong
  conclusions.

\subsection{Laplace-Runge-Lenz vector}

The Laplace-Runge-Lenz vector points towards the periapsis of the orbit 
from the center of motion, and therefore it would seem that
  observing this vector during a simulation would result in an
  immediate measure of the periastron advance. 
This vector
is defined in ADM coordinates in terms of the canonically conjugate 
position $\vec R$ and momentum $\vec P$ as~\cite{Damour-Schafer:1988}: 
\begin{equation}\label{eq:RungeLenz}
\vec A= \vec P \times \vec L - G M \mu^2 \frac{\vec R}{R}\,,
\end{equation}
where $\vec L = \vec R \times \vec P$ and $\mu$ is the reduced mass.
Unfortunately, the
magnitude of this vector is proportional to $e$, i.e., it will typically be very
small.  Moreover, it is computed as the difference between two large
terms that almost cancel each other, resulting in large numerical errors.
Furthermore, relativistic effects, such as gauge effects, might affect the two
 terms in Eq.~(\ref{eq:RungeLenz}) differently, thus disproportionately 
affecting the small difference $\vec A$. 
Yet another obstacle is that the
numerical data do not give the canonical position and momentum.  
For all these reasons, we found it impossible to measure the periastron advance
 from the Laplace-Runge-Lenz vector even for the binary run with $e\sim 0.05$.

\section{Discussion}
\label{sec:discussion}

We have dealt with three aspects of eccentricity in binary
  black hole simulations: how to measure eccentricity, its decay
  during the inspiral, and periastron advance.

With regard to techniques to measure eccentricity,
  this paper provides a
  systematic comparison between several different estimators.  The
  ones shown in
Figs.~\ref{fig:BCP},~\ref{fig:Jena},~\ref{fig:FilteredWavePhase} and
\ref{fig:FilteredWaveFrequency} each displayed a different behavior,
even though
these definitions reduce precisely to the usual eccentricity $e$ in
the Newtonian limit.  Differences appear mainly because the data
corresponds to a binary in the last phase of the inspiral before
merger when relativistic effects are significant---a regime
  in which the Newtonian relations between the orbital variables are
no longer valid.

The eccentricity estimator $e_{\rm BCP}$ (see Fig.~\ref{fig:BCP})
exhibits two very undesirable
  features: For the $e\sim 0.05$ simulation, $e_{\rm BCP}$ depends
  strongly on the choice of how separation between the black holes is
  measured (coordinate distance $D$ vs. proper separation $s$).
  For small eccentricities $e\sim 5\times 10^{-5}$, no regular
  oscillatory behavior is apparent, rendering $e_{\rm BCP}$ useless as an
  eccentricity estimator.  This might be 
  because it uses a definition where eccentricity comes in the 
next-to-leading term, and the leading order Newtonian expression is not 
satisfied. Also, the high power of the contribution of the orbital variable 
makes the eccentricity easily affected by high-order harmonic modes in the 
orbital variables. We have observed similar behavior when we explored
 alternative definitions of the 
eccentricity based on Newtonian formulas combining orbital variables. 

Eccentricity measures based on orbital quantities (see
Fig.~\ref{fig:Jena}) give the right amplitude (for $t<2500M$ in the
case of $e_\Omega$ and $e_D$), and the phasing is quite consistent
between the different eccentricity estimators.  For instance, the
orbital frequency is maximal when the separation is minimal.
However, for the low-eccentricity
  simulation ($e\sim 5\times 10^{-5}$) higher-order harmonics are
clearly visible as the binary approaches the merger, in particular 
for the coordinate separation $e_D$ and the
orbital frequency $e_\Omega$.  The eccentricity measured from the
proper horizon separation($e_s$) is affected least  by these coordinate
effects.

Eccentricity measures based on extracted gravitational waves
(see Figs.~\ref{fig:FilteredWavePhase} and
~\ref{fig:FilteredWaveFrequency}) 
  result in clean oscillatory behavior, even for eccentricities as
small as considered here. No high-order harmonics are noticeable in
the wave extrapolated to infinity during the time interval
considered.
The eccentricity is
calculated from the maximum and minimum values in the oscillating
function without concern for the coordinate location in the orbit. It
is especially straightforward to calculate numerically the
eccentricity from the wave phase extrapolated to infinity without
resorting to any notions of ``distance'' between the holes.
Computing eccentricity from the gravitational wave phase is
  therefore the preferred method.  Unfortunately, the gravitational
  wave phase is not as easily accessible as orbital quantities: One
  needs to extract gravitational waves, the waveform is delayed by the
  light-travel time to the extraction radius, and, for best results,
  one may have to extrapolate to infinity.  Therefore, in practice,
  eccentricity estimators based on orbital quantities may be useful
  for immediate diagnostics during a simulation, then confirmed and refined
  subsequently by eccentricity estimators based on gravitational wave
  properties.

Notice that the eccentricity measurement could be affected by noise sources 
such as the ``junk radiation'' early in the simulation or by poor boundary 
conditions causing radiation reflection at the outer boundary. These additional
oscillations could easily be interpreted as eccentricity. In principle,
however,
one should be able to distinguish them from the eccentricity by the frequency 
of the oscillation. 

The second part of this paper describes measuring the decay
  of orbital eccentricity during the inspiral of equal mass
  non-spinning black hole binaries, revisiting earlier
  work~\cite{Pfeiffer-Brown-etal:2007}.  For both simulations
  considered, we find that eccentricity measured via proper separation
  ($e_s$) and via gravitational wave frequency decays with the same
  power of proper of separation, $s^\beta$, with exponent
  $\beta\approx 1.4$. This is somewhat smaller than the value predicted by
  post-Newtonian expansions, $19/12\approx 1.58$.  The earlier work,
  which was based on fewer data-points at closer separation, found a
  distinctively smaller exponent when computing eccentricity from proper 
separation rather than from the orbital frequency.

The third part of this paper presents a measurement of periastron
advance for equal and unequal mass non-spinning black hole
binaries.  For eccentric binaries, periastron advance will result in a
characteristic modulation of the observed GW signal, and hence it is
important to quantify its frequency.  We find that the numerically
computed periastron advance $\Omega_\Phi/\Omega_r$ disagrees with both
3PN predictions for equal-mass binaries, as well as with the test-mass
limit of geodesic motion in a Schwarzschild background.  As shown in
Fig.~\ref{fig:precession}, the periastron advance for black hole
binaries lies roughly halfway between these two analytic calculations.
The unequal mass evolutions considered have very small eccentricities;
this is unfortunate for our current purposes, as this made it
impossible to measure periastron advance well enough to test reliably
the approach to the test-mass limit with increasing mass-ratio.
While the data appear to approach the test-mass limit as the
  mass ratio deviates from unity, cf. Fig.~\ref{fig:precession-qneq1},
  detailed confirmation  will have to await until this analysis is repeated
 with somewhat higher eccentricity runs in the future.  Nevertheless, even the
equal-mass case shows that periastron advance is yet another feature
of fully numerical calculations that is not accurately predicted by
post-Newtonian expansions.  To achieve agreement, one may have to go
to higher order post-Newtonian expansions, or one may have to incorporate finite-size
effects.  More pragmatically, for applications to gravitational wave
data-analysis, one might also introduce fitting parameters into the
post-Newtonian models, and choose these parameters to enhance
agreement with the numerical waveforms.

\begin{acknowledgments}
  We thank Geoffrey Lovelace for providing initial data for the
  large eccentricity run, and Luisa Buchman and Mark Scheel for providing the
  data for the unequal mass simulations.  
  Results obtained in this paper were produced using the Spectral Einstein
  Code (SpEC)~\cite{SpECwebsite}.
  This work is supported in
  part by grants from the Sherman Fairchild Foundation to Caltech and
  Cornell, and from the Brinson Foundation to Caltech; by NSF grants
  PHY-0601459, PHY-0652995, and DMS-0553302 at
  Caltech; by NSF grants PHY-0652952, DMS-0553677, PHY-0652929, and
  NASA grant NNX09AF96G at Cornell.  H.P. gratefully acknowledges
  support from the NSERC of Canada, from the Canada Research Chairs
  Program, and from the Canadian Institute for Advanced Research.
\end{acknowledgments}

\appendix

\section{PN periastron advance}

 In post-Newtonian approximations, the periastron advance was calculated to 3PN
order in~\cite{DJS2000} for circular orbits in terms of the frequency-related
parameter $x$. In the nonspinning circular case, the explicit expression for
$K$ is given by 
Eq.(5.11) of Ref.~\cite{DJS2000} in terms of the angular momentum density 
$j$ for circular orbits and the symmetric mass ratio $\nu\equiv m_1 m_2/(m_1 + m_2)^2$, where $m_1$ and $m_2$ are the masses of the two bodies, as
\begin{eqnarray}
K_{\rm circ} &=& \frac{3}{j^2}+\frac{1}{2}(45-12\nu)\frac{1}{j^4} 
                  +6\Big[\frac{135}{4}+(\frac{41}{64}\pi^2-\frac{101}{3})\nu
\nonumber \\
 &&   +\frac{53}{24}\nu^2 
      -\nu \omega_{\rm static} - \nu^2 \omega_{\rm kinetic}\Big]\frac{1}{j^6} \,,
\end{eqnarray}
where 
the value of the ambiguity parameter $\omega_{\rm static}$ was computed by 
Ref.~\cite{Damour01} to be zero, and the ambiguity parameter   
$\omega_{\rm kinetic}$ was shown to be $41/24$ by Ref.~\cite{Blanchet00a}.
The ratio $1/j^2$ is replaced for circular orbits by $1/j^2_{circ}$ where  
\begin{eqnarray}
\frac{1}{j_{circ}^2} &=& x \Big\{ 1- \frac{1}{3}(9+\nu)x+\frac{25}{4}\nu x^2
\nonumber \\ && 
-\frac{16}{3}\Big[\frac{1}{64}\left( 41\pi^2- \frac{5269}{6} \right)\nu + \frac{511}{192}\nu^2
 \nonumber \\ && 
-\frac{1}{432}\nu^3 -(\nu \omega_{\rm static} + \nu^2 \omega_{\rm kinetic}) \Big]x^3 \Big\}\,.
\end{eqnarray}

\section{Test-mass periastron advance for a Schwarzschild black hole}

 Test particles follow geodesics in the
  background spacetime, which here is given by the Schwarzschild metric:
\begin{equation}
ds^2 = -A^{-1}(r)\, dt^2 + A(r)\, dr^2 +r^2 d\Omega^2,
\end{equation}
where $A(r) = (1\! -\! 2 M/r)^{-1}$. From Ref.~\cite{Weinberg72}, these 
geodesic equations are given in term the radius $r$ of the position 
vector as a function of time $t$ by 
\begin{eqnarray}
\label{eq:geo1}
r^2 \frac{d\Phi}{dt} &=& J /A(r)\,,
\end{eqnarray}
and
\begin{eqnarray}
\label{eq:geo2}
A^2(r) \left( \frac{dr}{dt} \right)^2 +\frac{J^2}{r^2} -A(r) &=& -E 
\end{eqnarray}
where $E$ and $J$ are constants of motion.

Since we are interested in measuring the periastron advance, we obtain the shape of the
 orbit using 
Eqs.(\ref{eq:geo1}) and (\ref{eq:geo2}):
\begin{equation}
\label{eq:orbit}
\frac{A(r)}{r^4} \left( \frac{dr}{d\Phi} \right)^2 + \frac{1}{r^2} - \frac{A(r)}{J^2} = -\frac{E}{J^2}\,.
\end{equation}
At the perihelia and aphelia of a test particle bound in an orbit around a 
black hole of mass $M$, $r$ reaches its minimum $r_-$ and maximum $r_+$ when 
$dr/d\Phi$ vanishes, so we can write
\begin{equation}
\frac{1}{r_{\pm}^2} - \frac{A(r_{\pm})}{J^2 } = -\frac{E}{J^2}\,.
\end{equation}
From the above relation, the constants of motion $E$ and $J$ can be written
as
\begin{eqnarray}
E &=& \frac{A(r_+)r_+^2-A(r_-)r_-^2}{r_+^2-r_-^2}\,,
\end{eqnarray}
and
\begin{eqnarray}
J^2 &=& \frac{A(r_+)-A(r_-)}{1/r_+^2-1/r_-^2}\,.
\end{eqnarray} 
By integrating Eq.~(\ref{eq:orbit}), we find that the angle swept out by the position vector as $r$ increases from $r_-$ to $r_+$
is given by
\begin{eqnarray}
\label{eq:intprecession}
\Phi(r_+) &=& \Phi(r_-)  \\ \nonumber
 && + \int_{r_-}^{r_+} A^{1/2}(r) \left[ 
  \frac{A(r)}{J^2} - \frac{E}{J^2} - \frac{1}{r^2}
  \right]^{-1/2}\frac{dr}{r^2}\,.
\end{eqnarray}
Then the orbit precesses in each revolution by an angle $\Delta \Phi$ defined 
as
\begin{equation}
\Delta \Phi = 2|\Phi(r_+)-\Phi(r_-)| -2 \pi\,.
\end{equation}

To compute the periastron advance $K$ as a function 
of the orbital frequency $\Omega_\Phi$, 
we pick a set of values for $(r_-,r_+)$ such that $r_+=r_-+\epsilon$ where 
$\epsilon$ is a small positive number. The fractional periastron advance is 
estimated using Eq.~(\ref{eq:intprecession}), and the orbital frequency is 
estimated using Eq.~(\ref{eq:geo1}).
While $K$  can be computed using elliptic integrals for
Eq.(\ref{eq:intprecession}), in practice it is simpler to evaluate it by 
numerical quadrature.

\bibliography{References/References}

\end{document}